\documentclass{aa}
\input epsf.sty
\begin{document}

   \thesaurus{11     
              (11.03.1;  
               12.03.3;  
               12.04.1;  
               13.25.2  
               )}
   \title{Regularity in the X-ray surface brightness profiles of
galaxy clusters  and the $M$--$T$ relation.}


   \author{D.M. Neumann, M. Arnaud       }

   \offprints{D.M. Neumann}

   \institute{CEA/DSM/DAPNIA/Service d'Astrophysique, CEA-Saclay,\\
              L'Orme des Merisiers, B\^at. 709, F-91191 Gif-sur-Yvette,
              France\\
              email: ddon@cea.fr , arnaud@hep.saclay.cea.fr
}
\date{Submitted January 5, 1999} 

\titlerunning{Regularity in the cluster surface brightness profiles and the
$M$--$T$ relation}
\maketitle
%
%
\def \etal      {et al.\ }
\def \xray {\hbox{X--ray} }
\def \rhocrit {\hbox{$\rho_c$}}
\def \fbargas {\hbox{${\bar f}_{gas}$}}
\def \fgas {f_{\rm gas}}
\def \rc  {r_{\rm c}}
\def \rs  {r_{\rm s}}
\def \xc {x_{\rm c}}
\def \xs  {x_{\rm s}}
\def \rdc  {r_{\rm VT200}}
\def \Mdc  {M_{\rm VT200}}
\def \TX {T_{\rm X}}
\def \einstein {\hbox{\it Einstein }}
\def \einsteinmpc {\hbox{\it Einstein/MPC}}
\def \asca {\hbox{\it ASCA }}
\def \exosat {\hbox{\it EXOSAT }}
\def \rosat {\hbox{\it ROSAT }}
\def \ginga {\hbox{\it GINGA }}

 \def
\Mstar {\hbox{$M_{\star}$}} \def \LxT {\hbox{$L_X$--$T$} }
\def
\betamodel {\hbox{$\beta$--model} }
\def \betamodels  {\hbox{$\beta$--models} }
\def \mdot {\hbox{$\dot{M}$} }
\def \ma {\hbox{$^{\mbox{m}}$}}
\def \ya {\hbox{$^{\mbox{y}}$}}
\def \da {\hbox{$^{\mbox{d}}$}}
\def \ea {\hbox{$^{\mbox{e}}$}}
\def \ssimeq {\!  \simeq \!} 
\def \Ho {{\rm\
H_{o}}}


\begin{abstract}

We used archival ROSAT observations to investigate the X-ray surface 
brightness profiles of a sample of 26 clusters in the redshift range 
$0.04<z<0.06$.  For 15 of these clusters accurate temperatures 
(k$T_{\rm X} > 3.5$ keV) were available from the literature.  The 
scaled emission measure profiles look remarkably similar above $\sim 
0.2$ times the virial radius ($r_{\rm VT200}$).  On the other hand a 
large scatter is observed in the cluster core properties.  We fitted a 
\betamodel (with and without excising the central part) to all the 
ROSAT profiles to quantify the structural variations in the cluster 
population, unraveling a robust quadratic correlation between the 
core radius and the slope parameter $\beta$.  We quantified the shape 
of each gas density profile by the variation with radius of the 
logarithmic slope, $\alpha_{\rm n}$.  The bi-weight dispersion of 
$\alpha_{\rm n}$ among the clusters is less than $20 \%$ for any given
scaled radii above $x=0.2$.  There is a clear minimum spread at $x = 
0.3$, which is related to the existence of a correlation between core 
radius and $\beta$.  These ensemble properties are insensitive to the 
exact treatment of a possible central excess when fitting the 
profiles.  On the other hand the scatter is decreased when the radii 
are scaled to $r_{\rm VT200}$. 

The regularity we found in the gas profiles at $x> 0.2$ supports the 
existence of an universal underlying dark matter profile, as already 
predicted by theoretical works.  It suggests that non gravitational 
heating is negligible for clusters with temperature above $\sim 3.5$ 
keV. The very large scatter observed in the core properties favor 
scenario where Cooling Flows are periodically erased by merger events.  
Our results are consistent with the classical scaling relation between 
Mass and Temperature ($M \propto T^{3/2}~(1+z)^{-3/2}$).  Accordingly 
the spread in the reduced mass profiles derived from the hydrostatic 
isothermal \betamodel is small.

      \keywords{Galaxies: clusters: general --
                Cosmology: observations, dark matter --
                X-rays: general
               }
   \end{abstract}

\section{Introduction}

There is growing observational evidence that the physical properties
of galaxy clusters obey scaling relations.  Evrard (1997) showed that
most observed clusters of galaxies have a similar fraction of hot gas,
as compared total mass, $\fgas~\sim 0.060~h^{-3/2}$.  This fraction
sets a lower limit on the baryon fraction in these dark matter
dominated objects.  Furthermore, studies focusing on the relationship
between the X-ray luminosity of the hot gas (or intracluster medium,
hereafter ICM) and its X-ray temperature done by Markevitch (1998),
Allen \& Fabian (1998) and Arnaud \& Evrard (1998) have revealed a relatively
tight correlation between these two parameters.  Recently Mohr \&
Evrard (1997) found that there exists a strong correlation between the
X-ray isophotal radius and the X-ray temperature of the hot ICM. The
study of Hjorth, Oukbir \& van Kampen (1998), based on a sample of
clusters with good X-ray temperature estimates and mass inferred from
lensing observations, suggests that there is a tight mass--temperature
relation.

Such scaling laws are expected if clusters of galaxy form a
homologous population and are an indication of an underlying
structural regularity in the cluster population.  Hence their study
can give useful insight into the formation and evolution of clusters
and the underlying cosmological scenario.  In addition they can
provide an efficient way to estimate some cluster properties, which
are difficult to measure directly.  In particular the M--T relation is
of special interest, in view of the cosmological importance of
determining cluster masses and the possibility to measure accurately the
temperature with current and future X-ray telescopes.

Structural regularity is expected on theoretical grounds (Teyssier, 
Chi\`eze \& Alimi 1997).  Numerical simulations by Navarro, Frenk \& 
White (1996, 1997) indicate that CDM-halos with masses spanning 
several orders of magnitudes follow a universal density profile, whose 
shape is independent of mass or cosmology.  The corresponding density 
profile of the hot gas captured in these cluster halos can be 
reasonably well described by an isothermal \betamodel, as shown by 
Eke, Navarro \& Frenk (1998).  The X-ray surface brightness profile 
reads, for such a model:
\begin{equation}
S(\theta) = S_{\rm 0}~\left[1+\left(\theta/\theta_{\rm
c}\right)^2\right]^{-3\beta+1/2}
\end{equation}
which translates into the gas density distribution:
\begin{equation}
n_{\rm g}(r) = n_{\rm g0}~ \left[1+\left(r/\rc
\right)^2\right]^{-3\beta/2}
\label{equ:rho}
\end{equation}
The \betamodel is known since the early \einstein observations (Jones
\& Forman 1984) to provide a good fit to the X-ray profile of galaxy 
clusters.  However Makino, Sasaki \& Suto (1998) recently found that 
the gas density profiles deduced from simulations are too steep to 
match with the typical core radii $\rc$ derived from these X-ray 
observations.  Moreover it is well known that there are large 
variations, from cluster to cluster, in the observed \betamodel shape 
parameters, $\beta$ and $\rc$.

In this paper we want to check the regularity of the gas distribution
in clusters, described by a \betamodel.  We selected for this study a
sample of nearby clusters observed with \rosat and we unravel a
relation between the core radius and the slope parameter $\beta$.  We
discuss the implications of our results for the variations of the gas
density profile shape and the existence of a mass--temperature
relation.

The paper is organized in the following way: we describe in
Sect.~\ref{sec:observ} our sample of clusters of galaxies and the
observations.  Sect.~\ref{sec:surface} shows the surface brightness
profiles of the clusters.  In Sect.~\ref{sec:isofit} we present our
isothermal $\beta$-fits.  Sect.~\ref{sec:corr} shows the correlation
between $\rc$ and $\beta$.  In Sect.~\ref{sec:density} we describe the
consequences of the correlation on the shape of the gas density
profiles and quantify its variations and in Sect.~\ref{sec:mass} we
study the mass profile and the $M$--$T$ relation.  In
Sect.~\ref{sec:conclusion} we discuss our results and give our conclusions.

Throughout the paper we assume $\Ho=50$km/sec/Mpc, $\Lambda=0$, 
$\Omega =1$ ($q_{\rm 0}=0.5$).

\section{X-ray observations}
\label{sec:observ}

\begin{figure}[t]
\epsfxsize=8.5cm 
\epsfbox{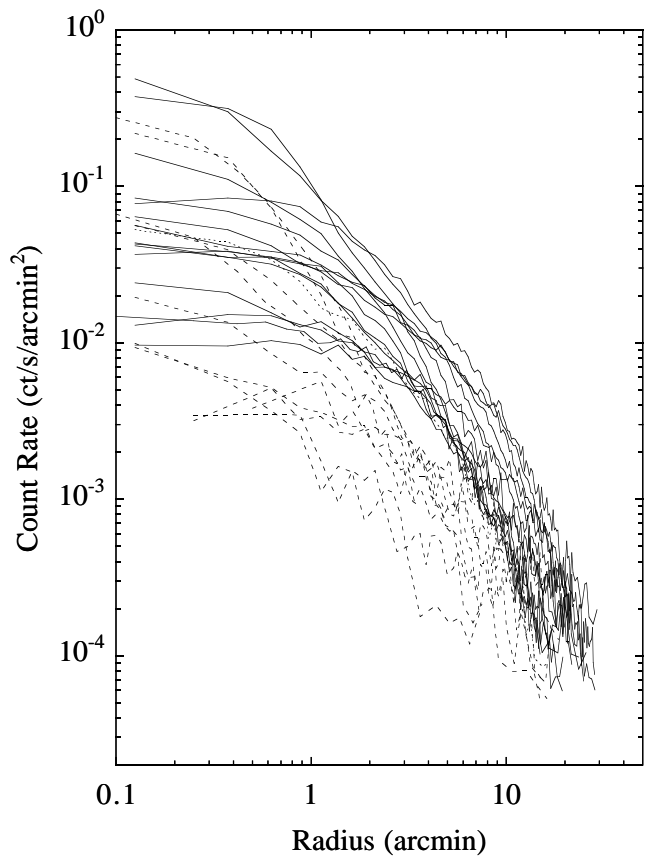} \caption{The ROSAT 
X-ray surface brightness profiles of the clusters in our sample.  The 
profiles are corrected for vignetting effects and background 
subtracted.  Full line: clusters in the spectroscopic subsample.}
\label{fig:sx}
\end{figure}

\begin{figure}[t]
\epsfxsize= 8.5cm \epsfbox{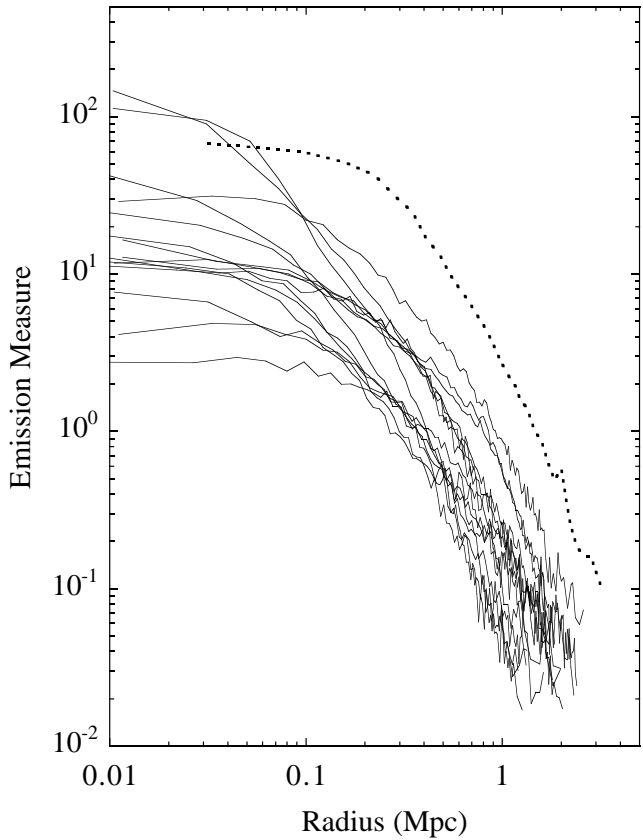} \caption{ The derived emission 
measure profile of the clusters in the spectroscopic sub sample.  The 
dotted line shows, for comparison, the emission measure profile of 
Abell 2163 (z=0.201, kT=14.6 keV, Elbaz, Arnaud \& B\"ohringer 1995) }
\label{fig:em}
\end{figure}

\begin{figure}[t]
\epsfxsize= 8.5cm \epsfbox{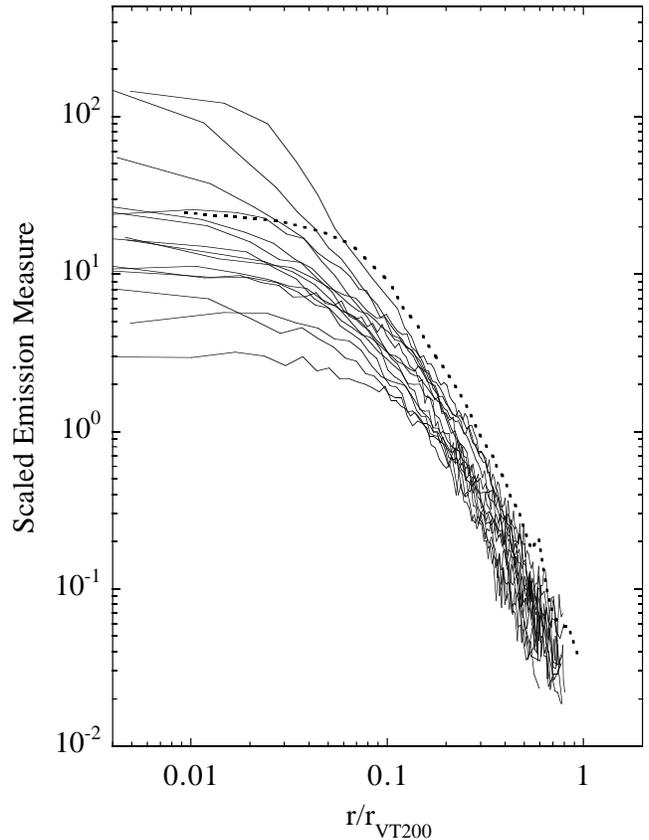} \caption{ The scaled emission
measure profile (Eq.~\ref{equ:emsc}) of the clusters in the
spectroscopic sub sample.  The radius is normalised to $\rdc$
(Eq.~\ref{equ:rvt200}).  Dotted line: Abell 2163 profile.  Beyond
$r/\rdc \sim 0.1$ the profiles look remarkably similar.}
\label{fig:emsc}
\end{figure}

We used X-ray imaging data retrieved from the ROSAT data base at MPE.  
The sample we built consists of
clusters of galaxies found by Abell, Corwin \& Olowin (1989) in the 
redshift range z=0.04-0.06, which were in the field of view of a ROSAT 
pointed observation with an off-axis angle less than 10 arcmin.  The 
redshift range is small enough to avoid to look at major evolutionary 
effects within the sample and the typical cluster size for these 
redshifts is well matched to the PSPC central field of view - mostly 
inside the rib structure of the instrument - or to the HRI field of 
view.  Furthermore we only selected clusters whose extended emission 
can be fitted with an isothermal $\beta$-model, i.e.  the cluster must 
show a clear center in the X-ray emission.  Finally the signal to 
noise ratio of the cluster must be high enough to allow accurate 
modeling of the data.

The list of the 26 clusters selected is given in
Tab.\ref{tab:betaresult}, as well as the exposure times and the
detector used.  The sample covers a large variety of morphological
types, from relaxed spherical symmetric clusters like A2107, to
clusters with substructures like A754 or A3559 and contains
non-cooling flow clusters (e.g. A119) as as well clusters known to
have strong cooling flows (e.g.  A85).

For the PSPC data we only used photons in the band 0.5-2.0~keV, for
the HRI we only took into account channel 2-9, in order to optimise
the signal-to-noise ratio.  For each data set we calculated the
exposure map, using the software implemented in the X-ray software
package {\it EXSAS} developed at MPE for the PSPC data and the
software developed by S. Snowden (Snowden 1998) for the HRI data.

We searched the literature for the emission-weighted mean temperature
of the clusters in our sample.  We only considered temperature
obtained with wide energy band experiments ({\it Einstein}/MPC, {\it
EXOSAT}, {\it GINGA} or {\it ASCA}) to avoid systematic errors and
exclude temperature estimates with statistical errors greater than
$50\%$.  The adopted temperature values and corresponding references
are given in Tab.~\ref{tab:betaresult} for 15 clusters, which
constitute our spectroscopic sub-sample.  Note that this sub-sample
does not contain low temperature clusters (${\rm k}T > 3.5 {\rm
keV}$), because only the brightest clusters of our sample have good
temperature measurements (see Fig.~\ref{fig:sx}).  All these clusters
have flux greater than $1.7~10^{-11} {\rm ergs/s/cm^2}$.  The
spectroscopic sample is $80\%$ complete at that flux limit, when
compared to the complete sample derived by Ebeling \etal (1996) from
the ROSAT All-Sky Survey.

\section{X-ray surface brightness profiles}
\label{sec:surface}

\begin{table*}
\caption[]{ROSAT observation analysis summary.  p/h stands for PSPC or
HRI. All quoted errors are $2-\sigma$ errors.  The redshifts are taken
from NED. When the best fit \betamodel A and B are identical, we only
give the parameters once.}
\label{tab:betaresult}
\begin{tabular}{l|ccc|cccr|ccrcl}
\hline
\multicolumn{1}{c}{}
& \multicolumn{3}{c}{model A} & \multicolumn{4}{c}{model B} & & & & & \\
\hline
Abell & $\beta$  & $\rc$ & $\chi^2_{\rm red}$ & $R_{\rm cut}$ &
$\beta$ & $\rc$ & $\chi^2_{\rm red}$ & $z$ & kpc/ & exp. & det. & kT \\
name &      & [kpc] & & [arcmin] &   & [kpc] & &  &
arcmin &
[ksec] & & in keV\\
\hline
76 & $0.60_{-0.15}^{+0.40}$ & $413_{-186}^{+383}$& 1.20 &    &
        &                   &    &0.042&68&1.7 & p & \\
85 & $0.53_{-0.01}^{+0.05}$ & $69_{-4}^{+4}$& 8.26 & 3 &
$0.70_{-0.03}^{+0.02}$ & $282_{-33}^{+32}$&1.63&0.052&83&15.9&p &
$6.1\pm0.2$ \ma\\
119 & $0.64_{-0.03}^{+0.04}$ & $463_{-40}^{+43}$& 1.44 &   &
             &                   &    &0.044&71&15.2&p & $5.8\pm0.6$ \ma\\
548 & $0.52_{-0.04}^{+0.06}$ & $132_{-33}^{+42}$& 1.35 & &
         &                   &    &0.042 & 68 & 10.9 & p & \\
754 & $1.04_{-0.04}^{+0.08}$ & $731_{-30}^{+32}$& 3.29 & 4 &
 $ 0.94_{-0.06}^{+0.07}$ & $853_{-67}^{+71}$&2.03&0.053&85&16.8&p &
$7.6\pm0.3$ $^{\mbox{a}}$\\
780 & $0.60_{-0.01}^{+0.01}$ & $62_{-2}^{+2}$& 4.75 & 2.25 &
$0.70_{-0.03}^{+0.02}$ & $148_{-23}^{+18}$&1.82&0.052&83&18.4&p &
$3.8\pm0.2$ \ma\\
S1101 & $0.57_{-0.01}^{+0.01}$ & $37_{-2}^{+3}$& 1.61 & 0.5 &
$0.59_{-0.02}^{+0.02}$ & $49_{-5}^{+5}$&1.37&0.058&92&47.4&h & \\
1991 & $0.59_{-0.02}^{+0.02}$ & $43_{-6}^{+6}$& 1.32 & 1&
$0.71_{-0.09}^{+0.13}$ & $142_{-49}^{+63}$&1.20&0.059&93&4.0 &p & \\
2107 & $0.60_{-0.02}^{+0.02}$ & $125_{-17}^{+18}$& 1.72 &2.75&
$0.76_{-0.08}^{+0.12}$ & $273_{-63}^{+74}$&1.11&0.041&66&8.3 &p & \\
2319 & $0.53_{-0.01}^{+0.05}$ & $208_{-15}^{+18}$& 1.86 &3.5&
$0.69_{-0.05}^{+0.06}$ & $496_{-73}^{+81}$&1.01&0.056&89&4.6 &p &
$9.1\pm0.2$ \ya\\
2589 & $0.60_{-0.02}^{+0.02}$ & $118_{-11}^{+12}$& 1.45 & 1 &
$0.62_{-0.03}^{+0.03}$ & $135_{-18}^{+19}$&1.36&0.042&68&7.3 &p &
$3.7^{+2.2}_{-1.2}$ \da \\
2626 & $0.54_{0.04}^{+0.06}$ & $54_{-13}^{+19}$& 3.55 &  &
                      &                   &    &0.057&90&27.8&h & \\
2657 & $0.54_{-0.01}^{+0.01}$ & $112_{-9}^{+7}$& 4.45 & 4&
$1.24_{-0.17}^{+0.27}$ & $689_{-93}^{+137}$&1.76&0.040&65&18.9&p  &
$3.7\pm0.3$ \ma\\
2717 & $0.57_{-0.02}^{+0.03}$ & $86_{-14}^{+17}$& 1.57 & 1 &
 $0.66_{-0.05}^{+0.07}$ & $150_{-30}^{+36}$&1.31&0.050&80&9.9 &p &
 \\
3093 & $0.59_{-0.10}^{+0.35}$ & $120_{-72}^{+153}$& 1.74&  &
                      &                   &    &0.058&92&8.1 &p & \\
3158 & $0.67_{-0.04}^{+0.04}$ & $274_{-30}^{+33}$& 1.15 & 1 &
$0.68_{-0.04}^{+0.04}$ & $289_{-33}^{+39}$&1.14&0.059&93&3.0 &p &
$5.5\pm0.6$ \da\\
3223 & $1.00_{-0.23}^{+0.45}$ & $842_{-224}^{+355}$& 1.53 &  &
                      &                   &    &0.060&95&7.7 &p & \\
3266 & $0.88_{-0.04}^{+0.05}$ & $688_{-45}^{+50}$& 5.43 & 3.5 &
$1.39_{-0.17}^{+0.19}$ & $1181_{-130}^{+140}$&1.91&0.059&93&13.6&p &
$7.7\pm0.8$ \ma\\
3301 & $0.63_{-0.05}^{+0.08}$ & $267_{-50}^{+63}$& 1.90 & 1.5 &
$0.73_{-0.07}^{+0.14}$ & $377_{-83}^{+108}$&1.70&0.054&86&8.9 &p& \\
3391 & $0.54_{-0.03}^{+0.04}$ & $221_{-33}^{+37}$& 1.35 & 1.25 &
$0.57_{-0.04}^{+0.05}$ & $259_{-47}^{+53}$&1.29&0.053&85&6.6 &p &
$5.7\pm0.7$ \ma\\
3532 & $0.66_{-0.05}^{+0.05}$ & $301_{-41}^{+48}$& 1.46 &2.25&
 $0.80_{-0.09}^{+0.13}$ & $461_{-90}^{+102}$&1.32&0.056&89&8.6 &p &
$4.4\pm1.5$ \ea\\
3558 & $0.58_{-0.01}^{+0.01}$ & $228_{-10}^{+12}$& 9.81 &2.75&
$0.72_{-0.02}^{+0.03}$ & $405_{-23}^{+24}$&2.47&0.048&77&29.5&p &
$5.5\pm0.3$ \ma \\
3559 & $0.56_{-0.10}^{+0.37}$ & $154_{-78}^{+213}$& 1.28 &  &
                      &                   &    &0.047&76&8.1 &p & \\
3562 & $0.47_{-0.01}^{+0.01}$ & $102_{-9}^{+10}$& 2.12 & 4.25&
$0.53_{-0.04}^{+0.06}$ & $245_{-151}^{+122}$&1.85&0.050&80&20.2&p &
$3.8\pm0.9$ \da\\
3667 & $0.72_{-0.09}^{+0.13}$ & $360_{-56}^{+68}$& 1.49 &2.67&
$0.67_{-0.12}^{+0.19}$ & $310_{-100}^{+121}$&1.33&0.055&87&42.3&h &
$7.0\pm0.6$ \ma\\
4059 & $0.58_{-0.01}^{+0.02}$ & $90_{-8}^{+9}$& 1.42 &1.5 &
$0.64_{-0.03}^{+0.04}$ & $151_{-27}^{+28}$&1.14&0.046&74&5.4 &p &
$4.1\pm0.3$ \ma\\
\hline
\end{tabular}
\smallskip
{\ma: Markevitch 1998;$^{\mbox{a}}$: Arnaud \& Evrard 1998;
 \da: David et al. 1993; \ya: Yamashita
(priv.  comm.); \ea: Edge \& Stewart (priv.  comm.  quoted in White, 
Jones \& Forman 1997)}
\end{table*}

Fig.\ref{fig:sx} shows the vignetting corrected X-ray surface
brightness profiles of all the clusters in the sample.  We bin the
photons into concentric annuli centered on the maximum of the X-ray
emission.  For the PSPC we use a width of 15 arcseconds per annulus
and a total of 200 annuli.  For the HRI we use a width of 10
arcseconds per annulus and a total of only 100 annuli, due to the
smaller field of view of the HRI. We cut out serendipitous sources in
the field of view or cluster substructures, if they show up as a local 
maximum. The background was subtracted using
data in the outer part of the field of view.

The emission measure along the line of sight at radius r, $EM(r)$, can
be deduced from the X-ray surface brightness, $S(\theta)$:
\begin{equation}
EM(r) = \frac{4~\pi~(1+z)^{4}~S(\theta)}{\Lambda(T,z)}~~~;~~~r =
d_{\rm A}(z)~\theta
\label{equ:sem}
\end{equation}
where $\Lambda(T,z)$ is the emissivity in the ROSAT band, taking into
account the interstellar absorption and the instrument spectral
response, and $d_{\rm A}(z)$ is the angular distance at redshift z.
$\Lambda(T,z)$ depends, although weakly, on the cluster temperature
and redshift.  The emission measure is linked to the gas density
$n_{\rm g}$ by:
\begin{equation}
EM(r) = \int_{r}^{\infty} \frac{n_{\rm g}^{2}(R)~R dR}{\sqrt{R^{2}-r^{2}}}
\label{equ:em}
\end{equation}

The shape of the surface brightness profile is thus governed by the
form of the gas distribution, whereas its normalization depends also
on the cluster overall gas content.  If clusters formed a structurally
similar population, all the brightness profiles would appear as
parallel curves on Figure~\ref{fig:sx} (they would differ by a
translation in the log-log plane).  More precisely similarity means
that the clusters constitute a one parameter population: each cluster
is characterized by its physical dimension $a_{i}$, and the
distribution of any given physical quantity $Q$ is described by a
dimensionless function $\widetilde{Q}$, common to all clusters:
$Q_{i}(r) = Q(a_{i}) \widetilde{Q} (x)$, where $x$ is the scaled radius
$x = r/a_{i}$.  The normalization factors $Q(a_{i})$ define the
scaling relations between the global quantities (e.g mass, mean
temperature, luminosity \ldots) in the cluster population.

We first examined if our data were consistent with the similarity
expected in the simplest model.  In the spherical collapse model, the
overall density contrast $\delta$ of virialized objects is fixed and
is of the order of $200$ for $\Omega=1$, and the natural scaling
radius is the virial radius defined as the radius containing this
density contrast\footnote{Note that numerical experiments support that
definition of the virial radius: the edge of the virialized part of
clusters is indeed found at a density contrast $\ssimeq 200$}.  Here
the density contrast is $\delta = \bar{\rho}/\rho_{\rm c}$, where
$\rho_{\rm c}(z)$ is the current critical density of the Universe,
$\rho_{\rm c}(z) = 3\Ho^2/8\pi G (1+z)^{3}$, and $\bar{\rho}$ is the
mean cluster density.  The virial radius is not directly measurable.
However, if, in addition, one assumes structural similarity, the
virial theorem provides a scaling relation between the virial mass
($\Mdc$) and radius ($\rdc$) and the overall X-ray temperature $T_{\rm
X}$: $\Mdc/\rdc \propto T_{\rm X}$, whereas by definition $\Mdc/(4/3
\pi \rho_{\rm c}(z) \rdc^{3}) = 200$.  This leads to the well known
scaling relations:
\begin{eqnarray}
\rdc & \propto & (1+z)^{-3/2}~\TX^{1/2} \label{rt} \\
\Mdc & \propto & (1+z)^{-3/2}~\TX^{3/2}
\label{mt}
\end{eqnarray}
The emission measure (Eq.~\ref{equ:em}) then scales as
\begin{equation}
EM(r) \propto \fgas^{2} \TX^{1/2} (1+z)^{9/2}\widetilde{EM}(x)
\label{equ:emsc}
\end{equation}
where $x=r/\rdc$ is the scaled radius, and $\widetilde{EM}$ is a
dimensionless function, the same for all clusters.  The gas mass
fraction $\fgas$ is not necessarily a constant but similarity implies
that it only depends on cluster mass (or equivalently temperature).

We thus considered the observed scaled emission measure profiles
$\widetilde{EM_{\rm X}}(x)$:
\begin{equation}
\widetilde{EM_{\rm X}}(x) =
\frac{S(\theta(x))}{(1+z)^{1/2}~\TX^{1/2}~\Lambda(T)}~~;~~ \theta(x)
= x~\frac{\rdc}{d_{\rm A}(z)}
\label{equ:emscx}
\end{equation}
defined from Eq.~\ref{equ:sem} and Eq.~\ref{equ:emsc} ignoring
possible variation of $\fgas$ with $\TX $.  To compute the scaling
radius $\rdc$ we used the normalisation factor obtained by Evrard,
Metzler \& Navarro (1996) from numerical simulations:
\begin{equation}
\rdc =
3690~\left(\frac{\TX}{10~\mbox{kev}}\right)^{1/2}(1+z)^{-3/2}~\mbox{kpc}
\label{equ:rvt200}
\end{equation}
Note that this normalization is simply for convenience, the exact calibration 
of the scaling relations does not matter to check similarity of the profiles.

The scaled emission measures profiles $\widetilde{EM_{\rm X}}(x)$ are
shown on Fig.~\ref{fig:emsc} for the clusters in the spectroscopic
sub-sample.  They can be compared to the corresponding unscaled
profiles $EM_{\rm X}(r)$ plotted on Fig.~\ref{fig:em}.  The scaled
profiles are clearly not identical, but several features are striking.
First, we detect the emission nearly up to the virial radius.  We are
effectively studying the large scale density distribution of the gas,
and not just the central cluster core.  Second, while the dispersion
is very large at low scaled radii ($\widetilde{EM_{\rm X}}$ at the
center span nearly 2 orders of magnitude), it rapidly decreases with
radius.  Beyond $r/\rdc \sim 0.1$ the profiles look remarkably similar
and agree to within about a factor of two.  We also note that the
scaling procedure has significantly decreased the difference between
the profiles, but only in that external region (compare
Fig.~\ref{fig:em} and Fig.~\ref{fig:emsc}).  The unscaled emission
measures show a relative standard deviation of about 96\% at 950~kpc
and 75\% at 1500~kpc.  In comparison, the relative standard deviation
of the scaled profiles is 41\% at $r/\rdc =0.3$ and 39\% at
$r/\rdc=0.5$, i.e the scaling has reduced the scatter by about a
factor of two.  As the temperature range of our spectroscopic
sub-sample is limited ($3.5 < {\rm k}T < 9.1{\rm keV}$), we plotted for
comparison the scaled profile of A2163.  It is one of the hottest cluster
known (Arnaud et al. 1992)
and is located at higher redshift than the clusters of our
sample.  It fits remarkably well within our sample.

In summary, the scaled emission measure profiles show indications for 
similarities in outer regions ($r > 0.2~r_{VT200}$).  On the other hand in 
the centers the profiles show a large dispersion, which we will 
discuss later on.  To further quantify the density profile shape and 
investigate its possible variations within the cluster population, we 
fit, in what follows, an isothermal \betamodel to each cluster 
profile.

\begin{table}
\caption{Description of the \betamodel}
\label{tab:fits}
\begin{tabular}{cl}
\hline
\betamodel & description \\
\hline
A & isothermal \betamodel fitted to the entire\\
& cluster surface brightness profile\\
\hline
B & isothermal \betamodel excising the central region\\
  & when an improvement of $\chi^2$ can be achieved\\
\hline
\end{tabular}
\end{table}

\section{Fitting	the	isothermal $\beta$-model}
\label{sec:isofit}

We performed two different \betamodel fits for each cluster.  First we
determined the \betamodel that fits at best the entire cluster
emission (hereafter \betamodel A).  However it is well known that the
overall 
\betamodel is a poor description of the central region of some
clusters where excess emission is observed (due to a cooling flow, a
central point source \ldots).  We indeed got in several cases very
large $\chi^{2}$ values (see Tab~\ref{tab:betaresult}).  Hence, we also
tried to minimise the reduced $\chi^2$ by excluding the central bins
from the fit.  The best fit \betamodel (hereafter \betamodel B) was
determined by increasing the radius of the cut-out region, $R_{\rm
cut}$, up to the radius where the reduced $\chi^{2}$ stopped decreasing.  We 
limited
$R_{\rm cut}$ to 350 kpc, as a too large excluded region in the center
precludes the
determination of the core radius.  Moreover it is safe to assume that
the contribution of any Cooling Flow excess is negligible beyond such
radius.

A brief description of \betamodel A and B is given in
Tab~\ref{tab:fits}.  The best fit parameters, $\beta$ and $\rc$, are
listed in Tab~\ref{tab:betaresult} for both models, together with the
$R_{\rm cut}$ values.  For about 1/4 of the clusters, excluding
central bins does not improve the fit.  In that case the best fit
\betamodels A and B are identical and we only give the parameters once
in the Table.

For the other clusters, excluding the central part yields on average
larger $\beta$ and core radii values (see Tab.~\ref{tab:betaresult}).
This is a well known effect.  If one tries to fit a \betamodel to the
entire profile when there is a central excess, too small core radius
and $\beta$ values are derived: decreasing $\rc$ allows us to fit
better the central part, while decreasing $\beta$ compensates for the
subsequent
flux deficit induced at large radii.  Although the \betamodel B
is better in that respect, it has also its drawback.  If the excluded
region is larger than strictly necessary to avoid the central excess,
the determination of the core radius is degraded: the uncertainty is
increased and the best fit value can be biased towards large values.
We will thus consider both models in the following.  They can be
viewed as two extreme \betamodels, allowing to assess the impact of the
exact treatment of the cluster core.  This is important since the core
properties vary greatly from cluster to cluster, as shown in the
previous section.

For A754, which is undergoing a merger phase, as indicated by many
authors (Zabludoff \& Zaritsky 1995; Henriksen \& Markevitch 1996;
Roettiger, Stone \& Mushotzky 1998), we varied the center used for the
concentric binning and adopted the profile that provides for the
global fit the lowest reduced $\chi^2$.

The most compact clusters A780 and A1991 were observed with the PSPC,
and their core image is somewhat blurred by the PSF ($\approx$
20-30~arcsec FWHM).  For A780 we estimated that the core radius is
overestimated by about 10\%, and for A1991 by around 20\%.
   For the PSPC pointing of A85 (the
cluster which has after A780 and A1991 the third smallest angular core
radius) the effect of the PSF is of the same order as the statistical
uncertainties.  For all other clusters, which either have a larger
core radius or were observed with the ROSAT HRI \footnote{the HRI has
a PSF with a FWHM of 4-5~arcseconds, which corresponds to about
4-8~kpc} we calculated that the effects of the PSF are much smaller
than the statistical errors.

All clusters with large error bars, such as A76, A3093, A3223 and
A3559, have clearly visible substructures.

\section{Correlation	between	the core radius and $\beta$}
\label{sec:corr}

\begin{figure}[t]
\epsfxsize=8.5cm \epsfbox{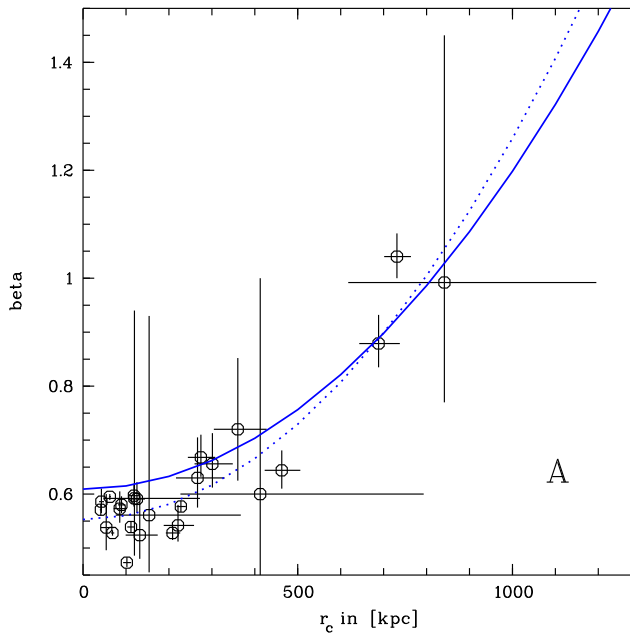} \caption{Slope parameter 
$\beta$ versus core radius $\rc$ for the 26 clusters of the sample 
(Tab.\ref{tab:betaresult}).  The isothermal \betamodel has been fitted 
to the entire ROSAT surface brightness profile of each cluster 
(\betamodel A).  The errors bars are $2$--$\sigma$ errors.  $\beta$ 
and $\rc$ are correlated.  The full line shows the best fit parabolic 
relation: $\beta = \beta_{\rm 
0}~\left[1+\left(\rc/\rs\right)^{2}\right]$, with $\beta_{\rm 0}=0.55$ 
and $\rs= 885$~kpc.  The dotted line shows the parabolic correlation 
obtained for the \betamodel B parameters: $\beta_{\rm 0}=0.61$ and 
$\rs= 1020$~kpc (see also Fig.~\ref{fig:corr_ocf}).}
\label{fig:corr_wcf}
\end{figure}

\begin{figure}[t]
\epsfxsize=8.5cm \epsfbox{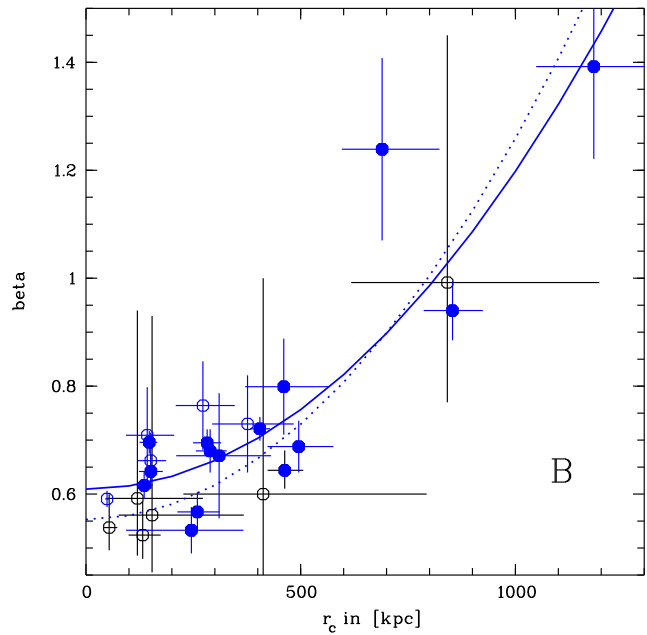} \caption{Same as 
Fig.~\ref{fig:corr_wcf} only that we now display the fit results of 
\betamodel B: isothermal \betamodel excising the central parts of the 
profile, if an improvement of reduced $\chi^{2}$ can be achieved (see 
Tab.\ref{tab:betaresult}).  Clusters belonging to the spectroscopic 
subsample are marked with filled circles.}
\label{fig:corr_ocf}
\end{figure}

\begin{figure}[t]
\epsfxsize=8.5cm \epsfbox{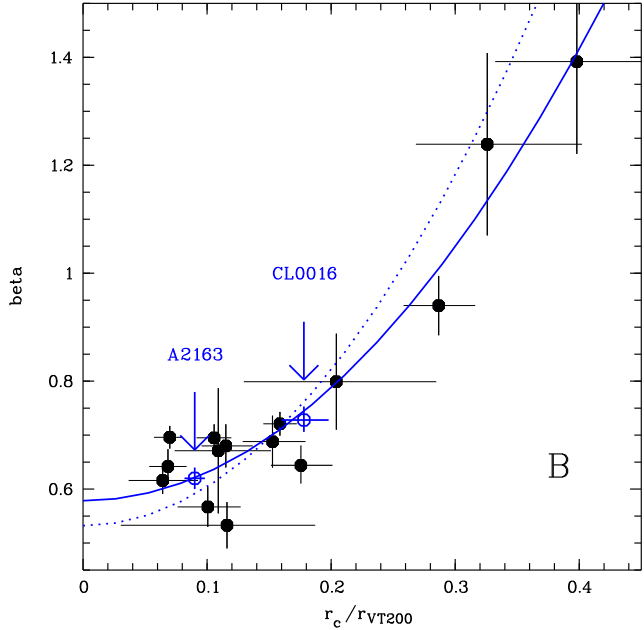} \caption{ The correlation 
between scaled core radius $\xc = \rc/\rdc$ and $\beta$ for the 15 
clusters of the spectroscopic subsample (see also 
Tab.\ref{tab:betaresult}).  The data points are for the \betamodel B 
and the error bars take into account the uncertainties in $\rdc$.  
Full line: $\beta$--$\xc$ relation for \betamodel B.  Dotted line: 
$\beta$--$\xc$ relation for \betamodel A.  We also display the results 
of A2163 located at $z=0.201$ (Arnaud \etal 1992; Elbaz, Arnaud \& 
B\"ohringer 1995) and CL0016 at $z=0.54$ (Hughes \& Birkinshaw 1998).  
}
\label{fig:corrtemp_ocf}
\end{figure}

\subsection{The $\beta$ and $\rc$ values and the significance of the
correlation}
\label{sec:var}

The slope parameter $\beta$ is not the same for all clusters, as would
be the case for perfectly similar \betamodel density profiles.  The
$\beta$ values span a wide range, $0.47-1.04$ and $0.52-1.4$ for the
\betamodel A and B, respectively.  The distribution is not symmetric,
it is clearly concentrated at low values, with rarer high values: the
median value is $0.59$ for A (0.68 for B), $\sim 85\%$ of the values fall 
in the
narrower range $0.47(0.52)-0.75$ and the standard deviation is only
$0.14(0.21)$ or $24\% (30\%)$ of the median value for model A (model
B).  The same general features are observed for the core radius
distribution.  The median values are $143$~kpc and $278$~kpc for
models A and B, respectively.

Fig.~\ref{fig:corr_wcf} displays the slope parameters $\beta$ against
core radius $\rc$, obtained with the \betamodel A.
Fig.~\ref{fig:corr_ocf} is the equivalent plot for \betamodel B. A
striking feature is that the data points are not distributed all over
the $\rc$--$\beta$ plane.  There is a clear trend of increasing slope
parameter with core radius, although the correlation is not very
tight, specially at low core radius values.  A quantitative measure of
the correlation significance was obtained from three different tests,
the Pearson's test, the Spearman's test, and the Kendall's $\tau$ test
(see Press \etal 1993).  All three tests (applied to fit A results) give 
a high correlation significance, higher than 99.97\%.  The
highest correlation significance is obtained with the Pearson's test,
with a value of $1.-2.99\times10^{-10}$.  The corresponding Pearson's
rank correlation is $r=0.902$.

\subsection{The form  of the $\rc$--$\beta$ relation}

The highest score obtained with the Pearson test indicates that the
correlation between $\rc$ and $\beta$ is close to being linear.  One
notes that the slope parameter is nearly constant at low core radii
and starts to increase with $\rc$ above $\rc \sim 250$kpc only,
which suggests a parabolic relation between the two quantities.

We  examined which relation, linear or parabolic, better accounts
for the data.  For conciseness, the following discussion is based on a
quantitative analysis of the parameters of the \betamodel A (similar
results are obtained with \betamodel B).

We considered the two parametric functions
  \begin{equation}
  \beta = \beta_{\rm 0}^l~\left[1+\left(\rc/\rs^l\right)\right]
  \label{equ:lin}
  \end{equation}
  and
  \begin{equation}
  \beta = \beta_{\rm 0}~\left[1+\left(\rc/\rs\right)^{2}\right]
  \label{equ:corr}
  \end{equation}
and fit them to the data.  Since both quantities $\beta$ and $\rc$ are
marred with errors and the uncertainties are highly correlated, there
is no well established method to define the best fit function
parameters and to quantify the goodness of the fit.  We used the
following empirical least-squares method.  Best fit parameters were
derived by omitting the errors on $\beta$ and $\rc$.  The $\chi^2$ was
minimised using the downhill simplex method (see Press \etal 1993).  We 
got $\beta_{\rm 0}^l = 0.49$ and $\rs^l = 870$~kpc for the linear 
function and $\beta_{\rm 0} = 0.55$ and $\rs = 885$~kpc for the 
parabolic function (the corresponding curve is plotted on 
Fig.\ref{fig:corr_wcf}).  To compare the goodness of the fits obtained 
with the two functions we performed a subsequent $\chi^2$-test 
including the error on the $\beta$ quantity only.  We utilised the 
$2-\sigma$ errors instead of the $1-\sigma$ errors.  As the 
uncertainties of $\beta$ and $\rc$ are correlated we assumed this 
could crudely take into account the two errors.

The reduced $\chi^2$ values for the entire cluster sample are 14.6 and
9.2 for the linear and parabolic function respectively.  The fit is
formally better for the parabolic relation but both $\chi^2$ are very
high.  If we exclude four clusters, A85, A780, A2319, and A3562, the
reduced $\chi^2$ drops to 1.68 for the parabolic function.  These four
clusters lie offset of the correlation and have extremely small
statistical error on $\beta$, less or equal to 2\%.  They boost
therefore the $\chi^2$ value.  The $\beta$-model parameters of
these four clusters change significantly when one cuts out the central
part (see Tab.\ref{tab:betaresult}) and it is likely that using the
global fit for these clusters is not a good approach.  On the opposite
we still obtain a reduced $\chi^2 $ of 3.36 with the linear relation
after the exclusion of the four principal outliers of this relation.
Only when excluding seven clusters, we do obtain a reduced $\chi^2$ of
1.78, which is still higher than the $\chi^2$ of the parabolic fit,
with four clusters excluded.

While keeping in mind that our criteria are not rigorous, we conclude
that a parabolic relation between $\rc$ and $\beta$ is a better
description of the correlation between these two quantities than a
linear relation.  Henceforth, we will only consider a parabolic
relation from now on.

\subsection {The best fit $\rc$--$\beta$ relation}
\label{sec:bestfit}

\begin{table}
\caption{The best fit parameters of the $\beta$--core
radius relations}
\label{tab:fitcrs}
\begin{tabular}{llll}
\hline
  Relation  & \betamodel & normalisation  & scale \\
\hline
$\beta$--$\rc$& A  &  $\beta_{\rm 0}=0.55$ & $\rs = 885$~kpc \\
$\beta$--$\rc$& B  &  $\beta_{\rm 0}=0.61$ & $\rs = 1020$~kpc \\
$\beta$--$\xc$& A  & $\beta_{\rm 0}^{\prime}=0.53$  &  $\xs = 0.27$   \\
$\beta$--$\xc$& B  & $\beta_{\rm 0}^{\prime}=0.59$  &  $\xs = 0.34$  \\
\hline
\end{tabular}
\end{table}

We determined the best fit parabolic relation $\rc$--$\beta$ for both
$\beta$--models A and B. The robustness of the $\rc$--$\beta$ relation
must be noted; it is not sensitive to the exact treatment of the
central part of the cluster profile.  From the \betamodel B data,
plotted on Fig.~\ref{fig:corr_ocf}, we got $\beta_{\rm 0} = 0.61$ and
$\rs = 1020$~kpc.  These values are close to those obtained for the
\betamodel A: $\beta_{\rm 0}$ is increased by only $\sim 10\%$ and
$\rs$ by $\sim 14\%$.  The reduced $\chi^2$ is smaller:
$\chi^2= 1.92$ for the complete cluster sample; the $\chi^2$ drops to
1.42 if one excludes A780.  Part of the improvement is certainly an
artifact due to the larger uncertainties on $\beta$.  The two
relations look very much alike, as can be seen on
Fig.~\ref{fig:corr_ocf}.

As $\beta$ is a dimensionless quantity, while $\rc$ is not, one would
rather expect a relation between $\beta$ and a {\it scaled} core
radius.  Fig.~\ref{fig:corrtemp_ocf} shows the correlation between
$\beta$ and the scaled core radius $\xc =\rc/\rdc$, where the virial
radius $\rdc$ is given by Eq.\ref{equ:rvt200} and the data points are
from the \betamodel B. We fitted the parabolic relation
\begin{equation}
\beta = \beta_{\rm 0}^{\prime}\left[1+\left(\xc/\xs\right)^2\right]
\label{equ:corrsc}
\end{equation}
to the cluster parameters of the 15 clusters in the spectroscopic
subsample.  We obtained $\beta_{\rm 0}^{\prime} = 0.59$ and $\xs =
0.34$ for the \betamodel B and similar results $\beta_{\rm 0}^{\prime}
= 0.53$ and $\xs = 0.27$ for the \betamodel A (see also
Tab.\ref{tab:fitcrs}).

An important question is whether $\beta$ is more tightly related to
$\rc$ or to $\xc$.  We cannot definitely answer this question with the
present spectroscopic subsample.  It is too small and covers a too
narrow range in temperature and redshift.  There is only a factor 2.4
between the minimum and maximum temperature.  As $\rdc$ only scales as
$\sqrt{\TX}$, the effect of the scaling cannot be dramatic.  However
there are several indications that the `scaled' relation is the most
relevant one, as expected.  First we compared
Fig.~\ref{fig:corrtemp_ocf} to Fig.~\ref{fig:corr_ocf} at large
$\beta$ values ($\beta > 0.65$), where the dependence of $\beta$ with
$\rc$ is maximal.  For the clusters in the spectroscopic subsample
(filled circles) the scatter around the best fit relation has been
decreased by the scaling process.  Second we considered test
clusters outside the T and z range of our sample.  A2163, the highest
temperature cluster, fits well with both the $\beta$--$\xc$ relation
and the $\beta$--$\rc$ relation, so this is not conclusive.  On the
other hand Cl0016+16, the most distant cluster with good estimates of
$\beta$ and $\rc$ ($\beta=0.728$, $\rc =298$~kpc) fits much better
with the $\beta$--$\xc$ (see Fig.~\ref{fig:corrtemp_ocf}).  This issue
is further discussed in Sect.~\ref{sec:density}.

A summary of all fits of the $\beta$-$\rc$ relation is given in
Tab.~\ref{tab:fitcrs}.  We do not give errors on the relation
parameters, $\beta_{0}$ and $\xs$ (or $\rs$).  We do not know of any
proper method to estimate them since the uncertainties on the related
quantities $\beta$ and $\rc$ are highly correlated.  We only suggest
that they could be of the order of the difference observed when
considering the two \betamodel A and B. We also want to stress that
the relation parameters we determine here are in very good agreement
with the direct study of the density profile shape presented below in
Sect.~\ref{sec:density}.

\begin{table}
\caption{The $\beta$--$\rc$ relation for various cluster samples}
\label{tab:relcheck}
\begin{tabular}{lllll}
\hline
Number of & Choice & \betamodel & $\beta_{0}$ & $\rs$ \\
clusters &criteria \\
 \hline
26 & none & A & $0.55$ & $885$~kpc \\
26 & none & B & $0.61$ & $1020$~kpc \\
\hline
13 &  $r_{det} > 1.45$Mpc &A & $0.54$ & $859$~kpc \\
13 &  $\chi^{2}   < 1.6$ &A  & $0.56$ & $980$~kpc \\
13 &  $r_{det} > 1.45$Mpc &B & $0.60$ & $1050$~kpc \\
15 &  $\chi^{2} < 1.5$   & B & $0.64$ & $1080$~kpc \\
\hline
\end{tabular}
\end{table}

\subsection{Checking on spurious effects}

Could the derived correlation be an artifact of our data analysis?
This question must be addressed since our results could in principle be
affected by i) a choice of an inadequate model ii) the fitting process
iii) systematic errors in the modeling of the instrument
characteristics.  We now discuss each point in turn.

\subsubsection{The model}
The \betamodel is only an approximation of the real gas density
profile.  Hence, the parameters derived may depend on the quality of
the data and the cluster region used for the fit.  We have already
mentioned that a central excess can yield artificially low $\beta$
and core radii.  At large radii the \betamodel tends to a simple power
law.  If instead the density profile continuously steepens, the
$\beta$ value obtained from the fit will increase with the outermost
radius used in the fit (see for instance the numerical simulations of
Navarro, Frenk \& White 1995) and as a result probably also the derived 
core radius.  Moreover the magnitude of these effects will depend on 
the accuracy of the \betamodel approximation, which is likely to vary 
from cluster to cluster.  All together this could induce a spurious 
correlation between the derived parameters, similar to the one 
observed,i.e.  an apparent increase of $\beta$ with $\rc$.

To see if this a serious issue we first investigated how the
$\beta$--$\rc$ relation is affected by the size of the region used for
the fit.  We can first compare \betamodels A and B. We also derived
the relation for the half sample of the 13 clusters with the largest
$r_{det}$, where $r_{det}$ is the maximum radius at which the X-ray
emission is detected with a significance greater than $3\sigma$.
Second we investigated if the relation is sensitive to the quality of
the fit obtained with the \betamodel.  For this purpose we selected
the clusters with the lowest reduced $\chi^{2}$ and derived the
corresponding $\beta$--$\rc$ relation.  The $\chi^{2}$ value is an
indicator of the accuracy of the \betamodel approximation (but also
depends on the quality of the data).

The parameters of the relations derived for the various cluster
sub-samples are given in Tab.~\ref{tab:relcheck}.  In all cases, a
parabolic $\beta$--$\rc$ relation well accounts for the data.
Furthermore for each model (A or B), the parameters $\beta_{\rm 0}$
and $\rs$ depend very weakly on the selection on $r_{\rm det}$
($0.3-3\%$ effect) or on the selection on the quality of the fit
($2-10\%$ effect).  The largest discrepancy is actually seen between
\betamodels A and B, i.e if one includes or not the central excess
region in the fit.  The effect is small as discussed in the previous
section.  From this observed robustness of the $\beta$--$\rc$ relation
we concluded that it is unlikely that it is an artifact due to the
imperfection of the \betamodel.

\subsubsection{The fitting process}

To check the validity of the fitting procedure, we construct
artificial clusters, which lie outside the correlation and have
exposure times and countrates similar to the ones in our sample.  We
added Poisson noise to the images and treated these artificial
clusters in the same way as our real data and fit the isothermal
$\beta$-model to them.  The intrinsic values for $\beta$ and $\rc$
always lie well within the $2-\sigma$-errors of the fit results of the
simulated clusters images.  Thus, the fit procedure is correct and 
could not introduce a spurious correlation.

\subsubsection{The instrument}
As already discussed above, only two clusters could be affected by the
finite instrument PSF, namely A780 and A1991.  For the others, the PSF
effects are of the same order than the statistical uncertainties or
smaller.  Excluding these two clusters from the fit changes the
results of $\beta_{\rm 0}$ and $\rs$ by less than $3\%$.

We also verified the exposure map calculation (or vignetting effects).
We used the observations of clusters, that were not included in our
sample because we found no prominent extended cluster emission.  This
is the case for A195 (observing title MKN 359) observed by the PSPC,
and A1213 observed by the HRI (observing title 1113+29).  For both
pointings we divided the image of the actual data by the exposure map.
There is no significant variation in the overall photon distribution
when excluding obvious sources: the countrate is flat over the whole
field of view of the corresponding detector.  This also validates our
implicit assumption of a flat background in the FOV.


\section{The shape of the gas density profiles}
\label{sec:density}

A perfect similarity between the density profiles is ruled out by the
variation of $\beta$ from cluster to cluster.  On the other hand the
correlation found between the two shape parameters of the density
profile indicates some structural regularity in the gas distribution,
that we study in this section.  For that purpose we use the density 
profiles derived from our \betamodel fits of the observed brightness 
surface profiles.

\subsection{Quantifying the shape: the slope profile}

The shape of the gas density distribution may be quantified by the
variation with radius of its logarithmic derivative. 
This derivative is also  important for the estimate of the mass profiles (see 
Sect.~\ref{sec:mass}).  For the \betamodel form, it reads:
\begin{equation}
 \alpha_{\rm n}(r) = - \frac{d \log n_{\rm g}(r)}{d \log r} = \frac{3~\beta
 }{\left[1 + \left(\rc/r\right)^2\right]}
\label{equ:alphar}
\end{equation}
or equivalently, in term of the scaled radial coordinates:
\begin{equation}
 \alpha_{\rm n}(x) = \frac{3~\beta }{\left[1 + \left(\xc/x\right)^2\right]}
\label{equ:alphax}
\end{equation}
We will refer to this function as the slope profile: $\alpha_{\rm n}(r)$
is the slope of the gas distribution in the log--log plane at radius
$r$.  It depends on the two shape parameters of the \betamodel: the
core radius and $\beta$.

\subsection{The slope profiles for a theoretical parabolic $\beta$ - 
core radius relation}

For a parabolic relation between $\beta$ and $\rc$
(Eq.~\ref{equ:corr}), the slope profile may be written as:
\begin{equation}
 \alpha_{\rm n}(r) = 3~\beta_{\rm 0}~\frac{\left[1 +
 \left(\rc/\rs\right)^2\right]} {\left[1 +
 \left(\rc/r\right)^2\right]}
\label{equ:alphacorr}
\end{equation}
Similarly for a parabolic relation between $\beta$ and the {\it
scaled} core radius:
\begin{equation}
 \alpha_{\rm n}(x) = 3~\beta_{\rm 0}^{\prime}~\frac{\left[1 +
 \left(\xc/\xs\right)^2\right]} {\left[1 +
 \left(\xc/x\right)^2\right]}
\label{equ:alphacorx}
\end{equation}

The physical implications of the parabolic relation, is
straightforward from these two equations: i) if $\beta$ is related to
$\rc$, there is a specific radius, $\rs$, where the slope of the gas
density profile is the same for all clusters, ii) if $\beta$ is
instead related to the {\it scaled} core radius, this occurs at a
specific {\it scaled} radius, $\xs$.

Let us first assume a relation between $\beta$ and $\rc$.  At the 
specific radius $\rs$ the slope is equal to $3~\beta_{\rm 0}$, where 
$\beta_{\rm 0}$ is the normalisation factor of the parabolic relation.  
At very large radii, $r~>>~\rc$, $\alpha_{\rm n}(r)$ tends to 
$3~\beta$ - the \betamodel is equivalent to a power law - and there is 
no impact of the $\beta$--$\rc$ relation.  At low radii, $\alpha_{\rm 
n}(r)$ roughly scales as $\left(\rc/r\right)^2$ and in view of the large 
dynamical range of core radii values, it can vary by more than an 
order of magnitude from cluster to cluster.  This is illustrated on 
Fig.~\ref{fig:alpha} where we have plotted with thick lines two 
extreme slope profiles corresponding to the best fit $\beta$--$\rc$ 
parabolic relation derived in Sect.~\ref{sec:bestfit} 
(Eq.~\ref{equ:alphacorr} with parameters from Tab.~\ref{tab:fitcrs}).  
The first one, $\alpha_{\rm n}^{(1)}(r)$, corresponds to the minimum 
allowed $\beta$ value: $\beta = \beta_{\rm 0}$ or $\rc = 0$.  In that 
case $\alpha_{\rm n}(r)$ keeps constant with radius: $\alpha_{\rm 
n}^{(1)}(r) = 3~\beta_{\rm 0}$.  The second one, 
$\alpha_{n}^{(2)}(r)$, corresponds to the maximum $\beta$ value 
observed in our sample.  The two curves cross at $\rs$.

The same properties hold for the $\beta$--$\xc$ relation, in terms of
the scaled radial coordinates.

\subsection{The observed slope profiles}

\begin{figure*}
 \epsfbox{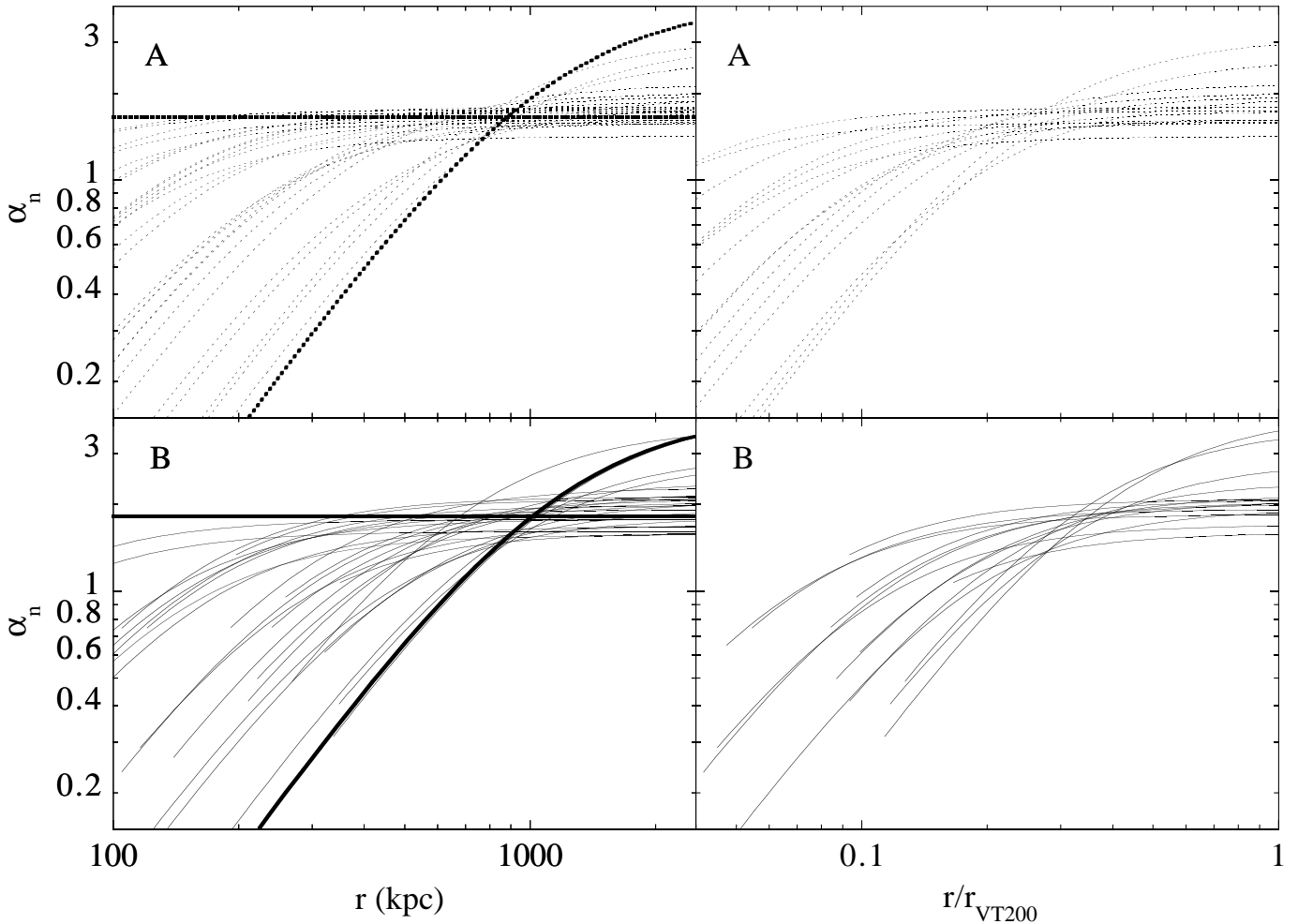} \caption{The variation with radius of the
 logarithmic slope of the gas density distribution.  Top-left
 panel:  the slope profiles $\alpha_{\rm n}(r) = - d \log (n_{\rm g}(r) /d \log 
 r$ versus radius $r$ derived from the best-fit \betamodel A, for all 
 clusters of the sample.  The bold lines correspond to the extreme 
 slope variations allowed by the $\beta$--$\rc$ correlation 
 (Eq.\ref{equ:corr} with parameters from Tab.~\ref{tab:fitcrs} and 
 Eq.\ref{equ:alphacorr}): the horizontal line is for 
 $\beta=\beta_{0}=0.55$, $\rc =0$ and the other line is for 
 $\beta=1.4$, $\rc=1100$~kpc. Top-right panel: the slope 
 profiles $\alpha_{\rm n}(x) =  - d \log (n_{g}(x) /d \log x$ versus scaled 
 radius $x =r/\rdc$ for the spectroscopic subsample (\betamodel A).  
   Bottom panels:  Same as top panels but for \betamodel B.  In that 
 case the profiles are plotted only beyond the excised central region.}
\label{fig:alpha}
\end{figure*}

\begin{figure*}
  \epsfbox{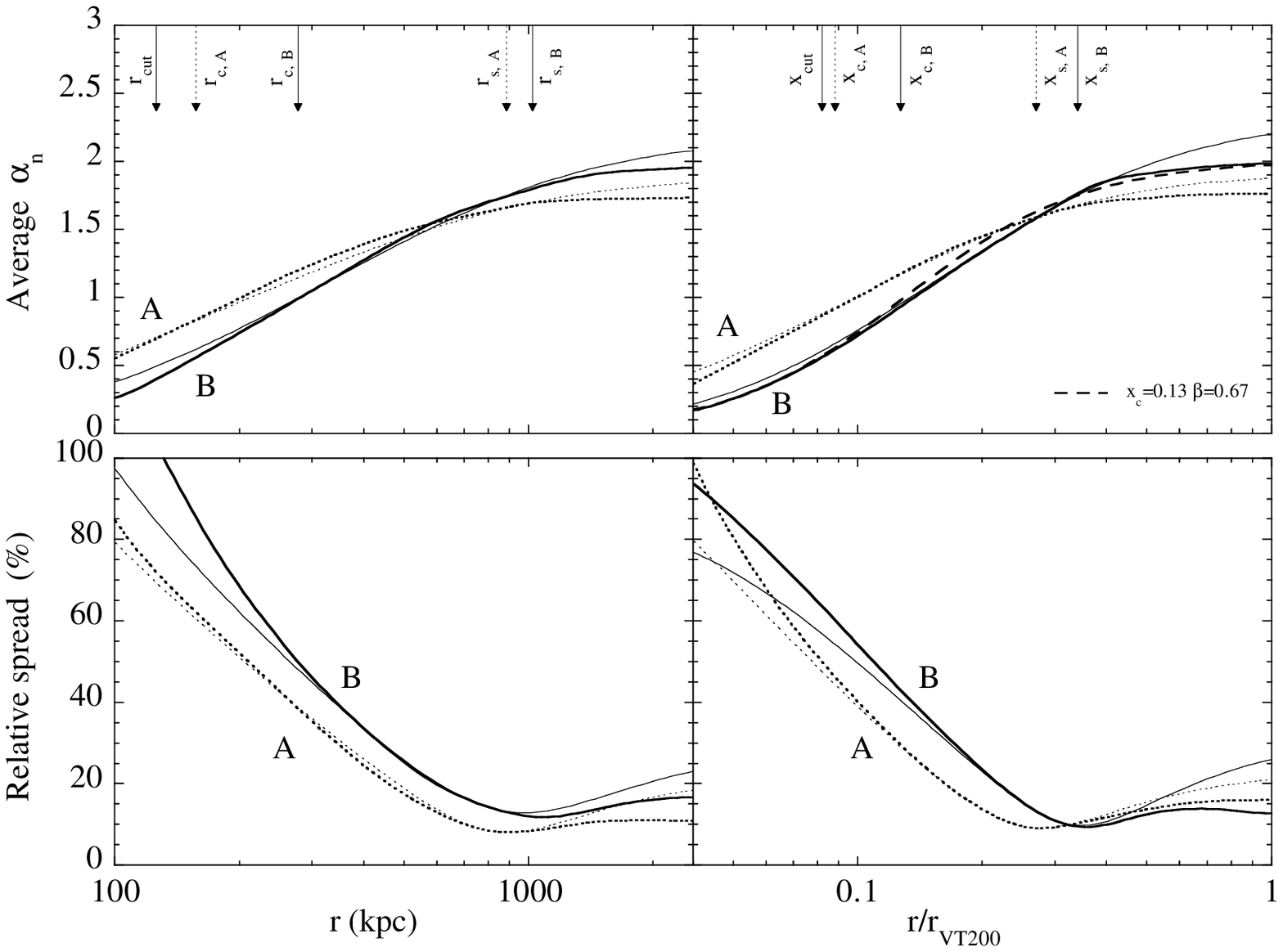} \caption{Structural variations within the
  cluster sample.  The average of $\alpha_{\rm n}$ (top panels) and its relative
  dispersion (bottom panels) are plotted as a function of radius
  (left) for the whole sample and scaled radius (right) for the
  spectroscopic subsample.  The relative dispersion is defined as
  spread normalised to the average value.    Thick lines: 
  bi-weight estimators.    Thin lines:  classical estimators. 
  Dotted lines:  results for \betamodel A.   Full lines: results
  for \betamodel B.  Arrows in left-top panel:  They indicate the
  bi-weight mean values of the core radii $r_{\rm c, A}$ and $r_{\rm
  c, B}$ for \betamodel A and B respectively; the bi-weight mean
  $R_{\rm cut}$; the specific radius $r_{\rm s,A} (r_{\rm s,B})$ of
  the $\beta$--$\rc$ correlation for \betamodel A(B).    Arrows in
  right-top panel:  Same for scaled coordinates.    Dashed line in
  top-right panel:  slope profile obtained for $\beta=0.67$,
  $\xc=0.13$.  }
  \label{fig:alphastat}
\end{figure*}

The slope profiles $\alpha_{\rm n}(r)$ derived from model A and B are plotted
on the left panels of Fig.~\ref{fig:alpha} for all clusters in our
sample.  On the right panels are plotted the profiles $\alpha_{\rm n}(x)$
against scaled radius for the spectroscopic subsample.

To quantify structural variations within the cluster sample, we
estimated at each radius, the center and spread of the distribution of
slopes at that radius.  We used both the classical estimators (mean
value and standard deviation) and the bi-weight estimators (bi-weight
location and scale).  The latter are more resistant and robust than
the former (Beers, Flynn \& Gebhardt 1990).  The results are shown in 
Fig.~\ref{fig:alphastat}.  The  ``average''  slope (top panels) and the 
``relative dispersion around the mean'', defined as the spread 
normalised to the average value (bottom panels) are plotted as a 
function of radius (left) for the whole sample and scaled radius 
(right) for the spectroscopic subsample.  The results of model A 
(dotted lines) and model B (full lines) are provided.  We also 
computed the bi-weight means of the core radii and cut-out radii 
$R_{cut}$, the locations of which are indicated by arrows in the figure.

\subsection{General properties}

The characteristics of the observed profiles are those expected from 
the best fit $\beta$--core radius relations we derived.  Let us 
consider for instance the slope profiles $\alpha_{\rm n}(r)$ derived 
from model B. At small radii the observed profiles lie well within the 
two extreme curves $\alpha_{\rm n}^{(1)}(r)$ and $\alpha_{\rm 
n}^{(2)}(r)$ corresponding to the best fit $\beta$--$\rc$ relation and 
the scatter is very large among the cluster population 
(Fig.~\ref{fig:alpha}, bottom-left panel).  This scatter diminishes 
with radius until it reaches a minimum at the specific radius $\rs = 
1020$ kpc (Fig.~\ref{fig:alpha} and Fig.~\ref{fig:alphastat} 
bottom-left panel).  At that radius the average $\alpha_{\rm n}$ value 
agrees well with the best fit $3~\beta_{\rm 0}$ value 
(Fig.~\ref{fig:alphastat} top--left panel).  The classical and 
bi-weight estimators give the same results, the slope distribution at 
that radius being close to Gaussian.  The residual scatter on 
$\alpha_{\rm n}$ at $\rs$ - the slope is not exactly the same for all 
clusters - can naturally be explained by the scatter around the best 
fit $\beta$--$\rc$ relation.  Above $\rs$ the scatter increases again, 
but saturate at a value of the same order as the variations of 
$3~\beta$, the asymptotic value of the slope.  For instance, at 2.5 
Mpc the bi-weight mean slope is 1.96 and the bi-weight relative spread 
is 17\%, very close to the corresponding values for
$3~\beta$: 1.98
and
18\% respectively.  At these large radii the bi-weight mean and
dispersion are smaller than the classical estimators, reflecting the
non-Gaussian $\beta$ distribution which presents a tail at high values
(see Sect~\ref{sec:var}).

These general properties hold both for \betamodel A and B. They do
not depend on the choice of the radial coordinates (physical or scaled
radius). However, there are some quantitative differences between the
corresponding data sets.

The relative dispersion of the slope $\alpha_{\rm n}(r)$ is higher for the
\betamodel B than for the \betamodel A at all radii.  This
may result in part from the larger uncertainties on $\beta$ and $\rc$,
which introduce extra scatter.  Using the scaled radii instead 
(Fig.~\ref{fig:alphastat} bottom-right panel), the  relative dispersion 
of $\alpha_{\rm n}$ for \betamodel B decreases.  The minimum 
dispersion, about 9\%, is similar for A and B in this case.
This further supports the idea that it is better to work with scaled 
coordinates, in agreement with the results of Sect.\ref{sec:bestfit}.  
Furthermore the bi-weight dispersion for model B now becomes slightly 
smaller than the dispersion for the model A, beyond the region where 
the dispersion is minimal ($x \sim 0.3$).  This is more satisfactory 
since the model B should be a better fit to the data at large radii as 
model A is expected to introduce extra scatter.  From now on we will 
only work in scaled coordinates.

The mean slope value is smaller at low radii (the density distribution
is flatter on average) and higher at high radii (steeper density
distribution) for the \betamodel B than for the \betamodel A
(Fig.~\ref{fig:alphastat} top panels).  This is a direct consequence
of the treatment of a possible central excess, as explained in details
in Sect.~\ref{sec:isofit}.  This central excess typically occurs
within $x_{cut} = 0.08$, the bi-weight mean cut-out radius.  Beyond
$r/\rdc\sim 0.1$ the systematic difference between the two models
becomes smaller than the intrinsic dispersion observed within the
cluster population.  The mean slopes for model A and B became 
identical in the region of minimum dispersion ($x \sim 0.3$).  This is 
consistent with the robustness of the $\beta$--$\xc$ relation already 
noted (Sect.\ref{sec:bestfit}).  This means that the determination of 
the mean slope around that scaled radius is very robust and does not 
depend on the exact model used to fit the data.

The mean slope profile can be approximated within a few percent with
the profile corresponding to the bi-weight mean core radius $\bar{\xc}
= 0.13 $ and $\beta$ value $\bar{\beta} = 0.67$ (see
Fig.~\ref{fig:alphastat} top-right panel).

\subsection{The standard density profile and structural variations in 
the cluster population}

In view of the shape characteristics of the observed density profiles,
it is natural to define a standard profile: this density profile
follows a \betamodel in scaled coordinates with $\xc = 0.13$ and
$\beta= 0.67$.  We recall that the cluster physical radius scales as
$\TX^{1/2}$.

The typical deviations from this standard profile are the following:
\begin{itemize}
\item A central excess  compared to the \betamodel can be observed in
the central part, typically within $x_{\rm cut} = 0.08$.  In that
region the scatter in density profile shape is very large (more than $100\%$
variation in logarithmic slope) among the cluster population.

\item Beyond $x = 0.2$ the bi-weight dispersion of the observed
 slopes $\alpha_{\rm n}(x)$ becomes less than $20\%$ of the slope of
the standard
\betamodel profile.

\item The dispersion is minimal at $x = 0.3$ where $\alpha_{\rm n}(x)
= 1.67 \pm 0.18$.  This corresponds to the observed correlation
between core radius and $\beta$ values.
\end{itemize}

Note that these values are derived from our spectroscopic-subsample
(Fig.~\ref{fig:alphastat}), which concerns relatively hot clusters.
Cooler clusters may present larger variations.

\section{Consequences for the total mass profile and the $M$--$T$ relation }
\label{sec:mass}

X-ray astronomers usually utilize the hydrostatic approach to
calculate the total mass of a cluster within a certain radius
\footnote{assuming that the cluster is in a dynamical relaxed state
and approximately spherically symmetric}:
\begin{equation}
 M_{\rm tot}(r) = -\frac{{\rm k}T}{\rm G \mu m_p}r\left( \frac{d \log n_{\rm
 g}(r)}{d \log r} + \frac{d \log T}{d \log r} \right)
\label{equ:mass1}
\end{equation}

\begin{figure}[t]
 \epsfxsize=8.5cm \epsfbox{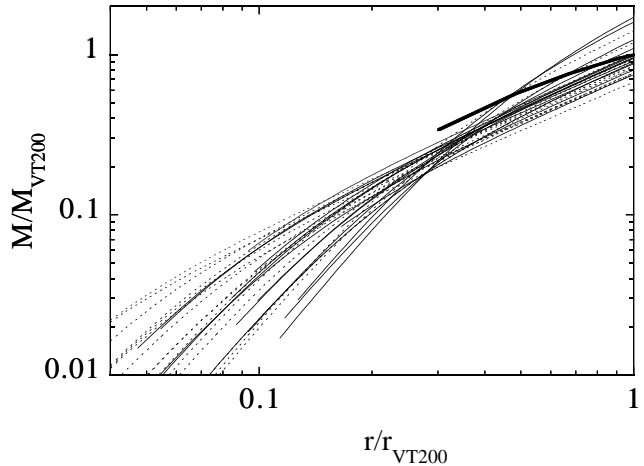} \caption{The normalised 
 mass profiles of the 15 clusters in the spectroscopic subsample.  For 
 each cluster the mass is derived from the hydrostatic equation and our 
 best fit isothermal \betamodel (Dotted line: \betamodel A; Full line: 
 \betamodel B).  The mass is normalised to $\Mdc$, the mass at density 
 contrast 200 derived from the numerical simulations of EMN96.  The 
 bold line is the profile derived by these authors.  }
\label{fig:massscAB}
\end{figure}

If the ICM is isothermal or if the temperature gradient can be
neglected in front of the density gradient, one may express the total
mass as a function of the X-ray temperature and the gas density
logarithmic slope $\alpha_{\rm n}$.  Using the scaled radial
coordinates one gets:
\begin{equation}
 M_{\rm tot}(x)=4.13\times 10^{15}\left(\frac{\TX}{10~\mbox{keV}}\right)^{3/2}
\frac{x ~\alpha_{\rm n}(x) }{(1+z)^{3/2}}~{\rm M_\odot}
\label{equ:mass2}
\end{equation}
where $\alpha_{\rm n}(x)$ is given by Eq.~\ref{equ:alphax} for a \betamodel.

It is convenient to normalize the mass profile to the virial mass at
density contrast 200 derived from numerical simulation (Evrard,
Metzler \& Navarro 1996):
\begin{equation}
 \Mdc = 2.92\times10^{15}~\left(\frac{\TX}{10~\mbox{keV}}\right)^{3/2}
 (1+z)^{-3/2}~{\rm M_\odot}
\label{equ:massvt}
\end{equation}
The corresponding scaled total mass profile, $\widetilde{M_{\rm tot}(x)}$, is
directly related to the variation of the gas density slope $\alpha_{\rm n}(x)$:
\begin{equation}
 \widetilde{M_{\rm tot}(x)} = \frac{M_{\rm tot}(x)}{\Mdc} =
 0.472~x~\alpha_{\rm n}(x)
\label{equ:masssc}
\end{equation}
Fig.~\ref{fig:massscAB} shows the total mass profiles,
$\widetilde{M_{\rm tot}(x)}$, derived from our best fit \betamodel A
(dotted lines) and B (full lines), for all the clusters of the
spectroscopic subsample.  As for $\alpha_{\rm n}(x)$, the differences
between the two models are smaller than the dispersion in the cluster
population.

The scatter in the total mass profiles is the same as for $\alpha_{\rm
n}(x)$ and is minimal at about $\xs=0.3$, the specific radius of the
$\beta$--core radius relation.  At that scaled radius $\alpha_{\rm
n}(\xs) = 1.67 \pm 0.18$ and $\widetilde{M_{\rm tot}(\xs)} = 0.236 \pm
0.025$.  In physical coordinates this corresponds to:
\begin{eqnarray}
 \rs & = &
 \frac{1.11}{(1+z)^{3/2}}
~\left(\frac{\TX}{10~\mbox{keV}}\right)^{1/2}~\mbox{Mpc}
 \\
 M_{\rm tot}(\rs) &=& \frac{(6.90 \pm
 0.74)\times 10^{14}}{(1+z)^{3/2}}
\left(\frac{\TX}{10~\mbox{keV}}\right)^{3/2}
{\rm M_\odot}
\label{equ:masskT}
\end{eqnarray}
Note that the numerical values in these equations do not depend of the
calibration of $ \rdc$ and $\Mdc$.  The radius of minimal dispersion
corresponds to a density contrast of $\delta = 1750$ (for the mass
computed from the isothermal \betamodel).

\begin{figure}[t]
 \epsfxsize=8.5cm \epsfbox{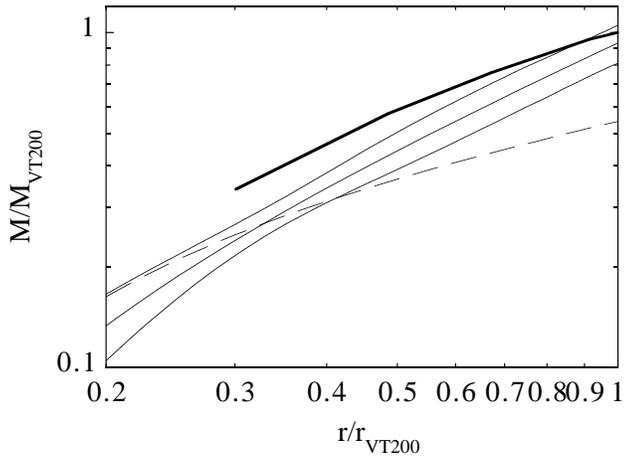} \caption{Comparison between 
 mass estimates for the spectroscopic subsample.  Bold line: Numerical 
 simulations of EMN96.  Full lines: The middle curve is the mean mass 
 profile for our sample derived from the isothermal \betamodel ($\xc = 
 0.13$ and $\beta= 0.67$); The top and bottom curves correspond to this 
 mean plus or minus the bi-weight dispersion.  Dotted line: The mean 
 mass profile for the same \betamodel but with a polytropic temperature 
 profile of index $\gamma= 1.2$, approximating the results of 
 Markevitch \etal (1998).  }
\label{fig:masscomp}
\end{figure}

The mean mass profile, corresponding to $\xc = 0.13$ and $\beta= 
0.67$, and the two curves corresponding to this mean plus or minus the 
bi-weight dispersion are plotted on Fig~\ref{fig:masscomp}.  For 
comparison we also plotted the mass profile derived from the numerical 
simulations of Evrard, Metzler \& Navarro (1996; hereafter EMN96; 
Table 5).  The isothermal \betamodel yields lower values but the 
discrepancy decreases with radius and is less than $10\%$ around $x =
1$.  Both models would give equal values at $ x=1$ for $\beta \sim 
\alpha_{\rm n}(1)/3 = 0.71$ (from Eq.~\ref{equ:masssc}), whereas the 
mean $\beta$ value is only $6\%$ lower.  \\

The isothermal assumption may give poor estimates of the total mass
if strong temperature gradients are present.  Recent observations with ASCA
suggest that the temperature does decrease with radius, although the
uncertainties on the temperature profiles are still large (Markevitch
\etal 1998).  To illustrate the impact of such a temperature gradient
we considered the composite projected temperature profile derived by
Markevitch
\etal (1998, Figure 7) from a sample of bright clusters.  This profile is
steeper than most of the profiles derived from available simulations
(Markevitch \etal 1998) and thus is likely to give an upper limit on
the influence of a temperature gradient, at least up to $x=0.5$ the
limit of the observations.  We found that the shape of the projected
temperature profile, $\bar{T}(x)$, can be well approximated by a
polytropic model with an index of $\gamma = 1.25$, a normalisation
factor $A=1.2$ and our mean
\betamodel density profile ($\beta=0.67$, $\xc = 0.13$):
\begin{equation}
\frac{\bar{T}(x)}{\TX} = A \left(\frac{n_{\rm g}}{n_{\rm
g0}}\right)^{\gamma -1}
= A \left[1 +
\left(x/\xc\right)^2\right]^{-\frac{3\beta}{2}\left(\gamma-1\right)}
 \end{equation}
Assuming that the observed temperatures correspond to the emission
weighted mean temperatures along the line-of-sight, it is easy to show
that the radial temperature profile, $T(x)$, also obeys the same
polytropic law with a higher normalisation factor, about $10\%$ higher
for the parameters
we derived.

For a polytropic gas, the temperature logarithmic slope $\alpha_{\rm
T}$, entering the hydrostatic equation (Eq.~\ref{equ:mass1}), is
simply related to the corresponding density slope:
\begin{equation}
\alpha_{\rm T}(x)
= (\gamma - 1)~\alpha_{\rm n}(x)
\end{equation}
 The corresponding mass profile,
plotted on Fig.~\ref{fig:masscomp}, is thus:
\begin{equation}
 \widetilde{M_{\rm tot}(x)} =
 0.472~\gamma \frac{T(x)}{\TX}~x~\alpha_{\rm n}(x)
 \end{equation}

The mass is larger than the isothermal estimate below $x \sim 0.35$ and
higher above. At $x =0.3$ the correction is only $4\%$, less than the
typical scatter. At large radii the differences are drastic: at $x=1$ the
isothermal estimate is 1.7 times higher than the estimate using the polytropic
model.  Whereas in the isothermal case we deduce a mass profile 
shallower than found in numerical simulation (EMN96), the mass profile is now 
slightly more peaked than the theoretical one. We must however emphasize 
that the temperature profile used is based on an extrapolation of the 
data beyond $x=0.5$.


\section{Discussion and Conclusions}
\label{sec:conclusion}

We have studied the shape of the surface brightness profiles of a
sample of clusters covering a large variety of morphological types.
The surface brightness distribution in the ROSAT band hardly 
depends on temperature and can be directly
translated into the gas density distribution.  Although the gas
density profiles of the clusters are not perfectly similar,  our quantitative
analysis gives evidence for structural regularity:
\begin{itemize}
\item
We found converging evidence that there is a scaling radius in
clusters which varies as $T^{1/2}$.  The surface brightness profiles
appear more similar once the physical radii are scaled.
The $\beta$--core radius relation is tightened and the scatter of the
density profile slopes is reduced.  However the small range in
temperature and redshift of our sample precludes any definitive
conclusion, in particular on the exact variation of the scale radius
with $T$.
\item
There is a parabolic correlation between the two shape parameters of
the \betamodel: the core radius and the slope.  Cluster density profiles
essentially constitute a one shape parameter family.
\item
We observe a very large scatter of the surface brightness profiles in 
the central part of clusters.  A central excess compared to the 
\betamodel can be present, typically within $x_{\rm cut} = 0.08$.  In 
this region the logarithmic slope can vary by more than a factor of 
$2$ and the emission measures at the center are spread over two orders 
of magnitude.
\item
Beyond a scaled radius of $x = 0.2$, the density profiles are 
remarkably similar in shape and resemble a \betamodel profile with a 
core radius of $\xc = 0.13$ and $\beta= 0.67$.  The bi-weight 
dispersion of the logarithmic slope $\alpha_{\rm n}(x)$ at a given 
radius is less than $20\%$.  At the virial radius the logarithmic
slope is close to $3\beta$.  Its distribution reflects the distribution 
of this parameter, which has a bi-weight dispersion of only $18\%$ but
presents a tail of rare high values.  The scatter is minimal at 
$x=0.3$ where $\alpha_{\rm n}(x)= 1.67 \pm 0.18$ with a gaussian 
distribution.  This minimum is related to the correlation between $\xc$ and 
$\beta$.  These results at $x=0.3$ are insensitive to the exact 
treatment of a possible central excess when fitting the profiles.
\end{itemize}

As the gas evolves in the potential of the dark matter it is unlikely 
that such regularity in the gas density profile shape can be observed 
without a comparable regularity in the dark matter profiles.  
Similarity of the dark matter profiles is naturally expected in 
hierarchical clustering scenario (Teyssier, Chi\`eze \& Alimi 1997; 
Navarro, Frenk \& White, 1997) and our results therefore give support 
to it.  However, the gas density is not the sole tracer of the 
gravitational potential.  If a universal profile for the dark matter 
is responsible for the similarity of the X-ray profiles, one also 
expects that the temperature profiles have similar shapes.  Both 
numerical simulations (e.g.  Navarro, Frenk \& White 1995; EMN96) and 
recent X-ray observations (Markevitch \etal 1998) indicate that this 
is probably the case.

Our results are consistent with a scaling radius which goes as 
$T^{1/2}$ as expected if clusters form at constant overdensity.  
However our sample is too small to exclude other scaling laws as, for 
instance, the slight variation of overdensity with mass obtained by 
Navarro, Frenk \& White (1997).  \\

Similarity breaking in the X-ray properties of clusters is expected if 
non adiabatic processes play an important role in the evolution of the 
gas.  These include radiative cooling and possible heating by galactic 
winds.

Radiative cooling is known to be more important in the cluster core 
where the cooling time, which scales as the gas density, is shorter.  
The presence of Cooling Flows of various strengths most probably 
explains the variation we observe in the density profiles below $x = 
0.1$.  The large scatter could be either due to intrinsic variation in 
the initial gas density in the core, and thus in the cooling time, or 
to variation in the Cooling Flow ``age''.  The first explanation is 
unlikely in view of the observed quasi-similarity of the scaled 
emission measure profiles at large radii.  Let consider perfectly 
similar clusters obeying the assumptions used in the scaling of the 
profiles.  The mean density of clusters in a given redshift range is a 
constant (formation at fixed overdensity) and since the gas mass 
fraction does not vary from cluster to cluster, the mean gas density 
is also a constant.  As the density profiles are similar in shape, the 
central density is also constant and so is the cooling time.  No major 
break of similarity in the core compared to external regions is thus 
expected, if clusters have the same ``age''.  The very large scatter 
observed in the core properties thus favors a scenario where Cooling 
Flows are recurrent phenomena that are periodically erased by strong 
mergers, a natural feature of hierarchical clustering (Fabian \etal 
1994).  In that case the observed variety of core properties in the 
cluster population would naturally reflect the statistics of the 
formation process via merger events.

Non gravitational heating, such as extra energy inputs by galactic 
winds, is thought to play an important role in the ICM physics.  In 
"pre-heating" scenarios, early winds add roughly a constant entropy to 
the system and thus affect cooler clusters more strongly, hence 
breaking the similarity law.  The main expected effect of these winds 
is to inflate the gas distribution.  Preheating scenarios can thus 
explain the observed steepening of the $L_{\rm X}$--$T$ and 
size-temperature relations relative to the similarity scaling (Arnaud 
\& Evrard 1998, Mohr \& Evrard 1997, Cavaliere, Menci \& Tozzi 1998).  
Further support to this scenario was recently given by Ponman, Cannon 
\& Navarro (1998) who showed that the entropy of cool systems is 
indeed higher than what can be achieved through gravity alone.  They 
consistently put into evidence similarity breaking in the surface 
brightness profiles of clusters: cooler clusters have systematically 
shallower profiles than hotter systems.  The regularity we found in 
the gas density profiles beyond the core does not contradict such 
results.  Our spectroscopic subsample comprises relatively hot 
clusters.  Our quantitative study simply shows that non gravitational 
heating is negligible above k$T \sim 3.5~{\rm keV}$.  This is in 
agreement with the quantitative deviations from similarity laws 
observed in clusters.  The gas entropy at $x=0.1$ plotted by Ponman, 
Cannon \& Navarro (1998) as a function of cluster temperature levels 
off from the similarity law only below $\sim 3$~keV. Similarly the 
variation of the structure parameter $\hat{Q}(T) = 
\langle{n_{gas}^{2}}\rangle/\langle{n_{gas}}\rangle^{2}$, which 
characterises the concentration of the gas distribution and plays a 
role in the steepening of the $L_{\rm X}$--$T$ relation, has been 
studied by Arnaud \& Evrard (1998, Figure 2).  An increase with 
temperature has been noted but the effect is again stronger below $4$ 
keV. The existence of such a "threshold" around $3-4$ keV clearly 
favors scenarios with constant entropy inputs.  In turn it shows that 
the outer regions of hot clusters are good tracers of purely 
gravitational processes.  \\

The mean gas density profile we derived is shallower than found in CDM 
simulations without winds.  For instance Navarro, Frenk \& White 
(1995, Table 2) found a mean core radius $\xc = 0.10$ and a mean 
$\beta$ of $0.82$, while Eke, Navarro \& Frenk (1998, Table 3) derived 
for a low $\Omega$ Universe average values of $\xc = 0.05$ and 
$\beta=0.73$.  If pre-heating is negligible in our spectroscopic 
subsample other effects can explain the difference.  First the 
transfer of entropy from the dark matter to the gas, which has been 
invoked to explain the segregation between these two components (Eke, 
Navarro \& Frenk 1998; see also Teyssier, Chi\`eze \& Alimi 1997), may 
be larger than predicted in the simulation.  We can also speculate 
that the dark matter distribution itself is actually less 
concentrated, as, for instance, expected if some hot dark matter is 
present (see also below).  \\

The gas structural variations throughout the cluster population are 
small beyond $x=0.2$ but depend on the radius considered.  The 
logarithmic slope $\alpha_{\rm n}(x)$ of the density profile presents a 
minimum spread at $x \sim 0.3$.  This scatter then increases with 
radius and some clusters present large asymptotic slopes compared to 
the mean asymptotic slope.  The \betamodel, used in our study, is 
reasonably accurate to estimate both the gas density distribution and 
total mass distribution, as shown by the hydrodynamic simulations of 
Schindler (1996) or EMN96.  However, by fitting such a model to simulated 
galaxy clusters, these last authors found that it introduces an 
extra-scatter in the total mass estimate and that this scatter 
increases with radius.  We probably see the same effect here.  In the 
hydrostatic isothermal \betamodel, the logarithmic slope of the 
density profile is directly linked to the total mass.  For a given 
dark matter profile, scatter in the gas density slope estimate can be 
introduced by i) errors in the measure due to the background 
determination and statistical errors; ii) departure from a spherically 
symmetric \betamodel for the gas distribution due to imperfection of 
the model, bimodality, ancient mergers, and high eccentricities; iii) 
departure from the hydrostatic equilibrium.  The first and third 
effects are expected to be larger in the outer regions.  On the 
contrary, the signal to noise is very high in the region around 
$x=0.3$ and the gas density slope, which is the derivative of the 
profile at that radius, is tightly constrained.  This probably 
explains why the scatter in the slope estimate is minimum there and 
why the derived slope value is insensitive to the exact model used to 
fit the data (model A or B).  In conclusion the observed variation of 
the scatter with radius is more likely an artifact due to the modeling 
of the data, rather than the direct consequence of similar variations 
in the dark matter profiles.

The slope at $x \sim 0.3$ is almost the same for all clusters.  Since, 
in the \betamodel, it depends on both $\beta$ and core radius, the 
values of these two parameters are correlated throughout the cluster 
population, as we did observe. This very correlation between $\beta$, 
that controls the outer slope, and $x_{c}$, that controls the core 
size, may indicate that the best model to fit the data is not the two parameter
\betamodel but some other model, involving only one shape parameter (scaling 
with the virial radius).  \\

As a result of the small scatter on $\alpha_{\rm n}(x)$, the spread in 
the scaled mass profiles derived from the hydrostatic isothermal 
\betamodel is small.  The uncertainty on the dark mass profile, 
deduced from the hydrostatic equation, is dominated by the systematic 
uncertainty on the temperature gradients, rather than by the intrinsic 
scatter in the gas density profiles.

The radius at which the scatter of the gas logarithmic slope is lowest 
is most likely in the region in which the \betamodel gives the most 
accurate result on the total mass of clusters.  It is also in the 
region in which temperature gradients are not expected to play a large 
role, as indicated both by numerical simulations (e.g.  EMN96) and by 
the observed temperature profiles (see previous section).  We thus 
propose the $M$--$T$ ($r$--$T$) relation, we derived at $x=0.3$ 
(Eq.~\ref{equ:masskT}) from the hydrostatic isothermal \betamodel, as 
a reference test point for numerical simulations.

In this relation the mass scales classically as 
$T^{3/2}~(1+z)^{-3/2}$.  The normalisation at $x=0.3$ is however $\sim 
30\%$ lower than found in the numerical simulations of EMN96, whereas
these authors did not observe any significant bias at that radius in 
the isothermal \betamodel estimate.  One can speculate again that the 
dark matter profile is in fact shallower than in CDM simulations, as 
expected if some HDM is present.  However the discrepancy could also 
indicate that the temperature spatial variations or bulk motions and 
 residual velocity dispersion of the gas are higher than expected.  
In that case the hydrostatic isothermal equation underestimates the 
real mass.  Large velocity fields are naturally associated with recent 
formation or accretion.  Therefore their properties should strongly 
depend on the specific history of each individual cluster and the 
corresponding velocity profiles are expected to present a high 
relative scatter.  If large kinetic energy is common in the cluster 
population, it would thus be surprising that the gas density profiles 
remain similar.  In conclusion we think that the regularity of the gas 
density profiles supports the validity of the hydrostatic equation.  
On the other hand quasi-similar temperature profiles corresponding to 
large temperature gradients could well be present.  It thus is essential to 
accurately measure the temperature profiles of large sample of clusters to 
establish firmly the normalisation of the $M$--$T$ relation and the 
shape of the dark matter profile.  This will be soon possible with XMM 
and AXAF.

\begin{acknowledgements}
We are very grateful for useful discussions with Hans 
B\"ohringer, Jean-Pierre Chieze and Alain Blanchard.  DMN is supported via an 
exchange program between the CNRS and the MPG.  This research has made 
use of the NASA/IPAC Extragalactic database (NED) which is operated by 
the Jet Propulsion Laboratory, California Institute of Technology, 
under contract with the National Aeronautics and Space Administration.

\end{acknowledgements}


\begin{thebibliography}{}

\bibitem[]{} Abell G.O., Corwin H.G. Jr., Olowin R.P., 1989, ApJS, 70, 1

\bibitem[]{} Allen S.W., Fabian A.C., 1998, MNRAS, 297, L57

\bibitem[]{} Arnaud M., Hughes J.P., Forman W., Jones C., Lachi\`eze-Rey M.,
Yamashita K., Hatsukade I., 1992, ApJ, 290, 345

\bibitem[]{} Arnaud M., Evrard A.E., 1998, MNRAS, astro-ph/9806353

\bibitem[]{} Beers T., Flynn K ., Gebhardt K., 1990, AJ, 100, 32

\bibitem[]{} Cavaliere A., Menci N., Tozzi P., 1998, MNRAS, astro-ph/9810498

\bibitem[ ] { }David L.P., Slyz A., Jones C., Forman W., Vrtilek
S.D., Arnaud K.A., 1993, ApJ, 412, 479.


\bibitem[ ] { } Ebeling H., Voges W., B\"ohringer H., Edge A.C.,
Huchra J.P., Briel J.G., 1996, MNRAS, 281, 799

\bibitem[]{}Eke V.R., Navarro J.F., Frenk C.S., 1998, ApJ, 563, 569

\bibitem[]{}Elbaz D., Arnaud M., B\"ohringer H., 1995, A\&A, 293, 337

\bibitem[]{} Evrard A.E., Metzler C.A., Navarro J.F., 1996, ApJ, 469, 494

\bibitem[]{} Evrard A.E., MNRAS, 1997, 292, 289

\bibitem[ ] { } Fabian A.C., Crawford C.S., Edge A.C., Mushotzky R.F.,
1994, MNRAS, 267, 779

\bibitem[]{} Henriksen M.J., Markevitch M., 1996, ApJ, 466L, 79

\bibitem[]{} Hughes J., Birkinshaw, M., 1998, ApJ, 501, 1

 \bibitem[]{} Hjorth J., Oukbir J., van Kampen E. 1998, MNRAS, 298, L1

\bibitem{} Jones C., Forman, W.,  1984, ApJ, 276, 38

\bibitem[]{} Makino N., Sasaki S., Suto Y., 1998, ApJ, 497, 555

\bibitem[]{} Markevitch M., 1998, ApJ , 504, 27

\bibitem[]{} Markevitch M., Forman W., Sarazin C., Vikhlinin A., 1998, ApJ , 503,
77

\bibitem[]{} Mohr J.J., Evrard A.E., 1997, ApJ, 491, 38

\bibitem[]{} Navarro J.F., Frenk C.S., White S.D.M., 1995, MNRAS, 275, 720

\bibitem[]{} Navarro J.F., Frenk C.S., White S.D.M., 1996, ApJ, 462, 563

\bibitem[]{} Navarro J.F., Frenk C.S., White S.D.M., 1997, ApJ, 490, 493

\bibitem[]{} Ponman T.J., Cannon D.B., Navarro J.F., Nature in press
 astro-ph/9810359 

\bibitem[]{} Press W.H., Teukolsky S.A., Vetterling W.T., Flannery B.P.,
1993, Numerical Recipes in C : The Art of Scientific Computing,
Cambridge University Press

\bibitem[]{} Roettiger K., Stone J.M., Mushotzky R.F., 1998, ApJ, 493, 62

\bibitem[]{} Schindler S., 1996, A\&A, 305, 756

\bibitem[]{} Snowden S., 1998, ApJ, in press

\bibitem[]{} Teyssier R., Chieze J.P., Alimi J.M., 1997, ApJ, 480, 36

\bibitem[]{} White, D., Jones C., Forman W., 1997, MNRAS, 292, 419

\bibitem[]{} Zabludoff A.I., Zaritsky D., 1995, ApJ, 447L, 21


\end{thebibliography}
\end{document}